\documentclass[10.5pt]{article}
\pdfoutput=1
\usepackage{amssymb,amsmath}
\usepackage{graphicx,float}
\usepackage{mathptmx}
\usepackage{graphicx}
\usepackage{times}
\usepackage{amsmath, amsthm, amssymb}
\usepackage{url}
\usepackage{subfig}
\usepackage{hyperref}
\usepackage{algorithm}
\usepackage{algorithmic}
\usepackage{color}

\usepackage{lscape}

\begin{document}
\title{The topology of higher-order complexes 
  associated with brain-function hubs in human connectomes}

\author{Miroslav Andjelkovi\'c$^{a,b}$, Bosiljka Tadi\'c$^{a,d}$,
  Roderick Melnik$^{c,e}$\\
\vspace{3pt}
{\small $^a$Department of Theoretical Physics, Jo\v zef Stefan Institute,
Ljubljana, Slovenia; }\\
{\small $^b$ Institute for Nuclear Sciences Vin\v ca,  University of Belgrade
11000 Belgrade, Serbia;}\\
{\small $^c$MS2Discovery Interdisciplinary Research
Institute, M2NeT Laboratory}\\
 {\small and Department of Mathematics, Wilfrid
Laurier University, Waterloo, ON, Canada;}\\  
{\small $^d$Complexity Science Hub,
Josefstaedter Strasse 39, Vienna, Austria;}\\
{\small $^e$BCAM - Basque Center for Applied Mathematics, Alameda de Mazarredo
14, E-48009 Bilbao, Spain}\\{\small  bosiljka.tadic@ijs.si,
http://www-f1.ijs.si/$\sim$tadic/ }} 

\vspace{3pt}
\date{ }
\maketitle
\tableofcontents
\newpage
\begin{abstract}
Higher-order connectivity in complex systems described by
simplexes of different orders provides a geometry for
simplex-based dynamical variables and interactions. Simplicial
complexes that constitute a functional geometry of the human connectome can be crucial for the brain complex dynamics.  In this context, the best-connected brain areas, designated as hub nodes,  play a central role in supporting integrated brain function. Here, we study the structure of simplicial complexes attached to eight global hubs in the female and male connectomes and identify the core networks among the affected brain regions. These eight hubs (Putamen, Caudate, Hippocampus and Thalamus-Proper in the left and right cerebral hemisphere) are the highest-ranking according to their topological dimension, defined as the number of simplexes of all orders in which the node participates. 
Furthermore,  we analyse the weight-dependent heterogeneity of
simplexes.  We demonstrate changes in the structure of identified core
networks and topological entropy when the threshold weight is
gradually increased. These results highlight the role of higher-order interactions in human brain networks and provide additional evidence for  (dis)similarity between the female and male connectomes.
\end{abstract}
 
\vspace{2pc}

\section{Introduction\label{sec:intro}}
Recent advances in the science of complex systems aim for a better
understanding of 
the higher-order connectivity as a possible basis for their
emerging properties and complex functions. Beyond the framework of
pairwise interactions, these connections described by
simplexes of different sizes provide the geometry
for higher-order interactions and  simplex-related dynamical variables. 
One line of research consists of modelling and analysis of the
structure of simplicial complexes in many complex systems, ranging from the
human connectome \cite{Brain_weSciRep2019} to quantum physics \cite{Geometries_QuantumPRE2015} and materials
science \cite{AT_Materials_Jap2016,we-SciRep2018}. Meanwhile,
considerable efforts aim at understanding the impact of geometry on
the dynamics. 
In this context, the research has been done on modelling of the
simplex-based synchronisation processes 
\cite{HOC_Synchro_ArenasPRL2019,HOC_synchroexpl_Bianconi2019}, on
studying the related spectral properties of the underlying networks
\cite{we-SpectraPRE2019,Spectra-rapisarda2019}, as well as on the interpretation of the dynamics of the brain \cite{HomologyBrain_petriJRSI2014,HOC_Brain_ReimanFrontCompNeuro2017,Brain_Hubs-dynamicsReview2018} and other complex
dynamical systems \cite{HOC_Kuehn2019}.

Recently, mapping the brain imaging data \cite{BrainMapping_Zhang2018} to
networks involved different types of signals across spatial and
temporal scales; consequently, a variety of structural and functional
networks have been obtained
\cite{BrainConnectivityNets_2010,BrainNetParcellation_Shen2010,EEGnets_ICAreconstr_PLOSone2016,Xbrain_hyperscanning}. 
This network  mapping enabled getting a new insight into the
functional organisation of the brain \cite{Brain_phys1review,Brain_phys2}, in particular,  based on the standard and deep graph theoretic methods
\cite{Sporns_BrainNets_review2013,brainGraph_metods2014,GD_Bud_PLOS2015} and the algebraic topology of graphs \cite{Brain_weSciRep2019,Xbrain_wePLOS2016}.
The type of network that we consider in this work is the whole-brain
network \textit{human connectome}; it is mapped from the \textit{fMRI}
data available from the human connectome project \cite{HCP_Neuroimage}, see Methods. The network nodes are identified as the grey-matter anatomical brain regions, while the edges consist of the white-matter fibres between them.  Beyond the pairwise connections, the human connectome
 exhibits a rich structure of simplicial complexes and short cycles
 between them, as it was shown in \cite{Brain_weSciRep2019}.
 Furthermore, on a mesoscopic scale, a typical structure with anatomical modules is observed.
It has been recognised
\cite{brainNets_tasksPNAS2015,Brain_modularNNeurosci2017} that every module has an autonomous function, which contributes to performing complex tasks of the brain. Meanwhile, the integration of this distributed activity and transferring of information between different modules is performed by very central nodes (hubs) as many studies suggest, see a recent review \cite{Brain_Hubs-dynamicsReview2018} and references therein. Formally, hubs are identified as a group of four or five nodes in each brain hemisphere that appear as top-ranking according to the number of connections or another graph-centrality measure. Almost all formal criteria give the same set of nodes, which are anatomically located
deep inside the brain, through which many neuronal pathways go.
Recently, there has been an increased interest in the research of the hubs
of the human connectome. The aim is to decipher their topological configuration and how they fulfil their complex dynamic functions. 
For example, it has been recognised that the brain hubs are mutually
connected such that they make a so-called ``rich
club'' structure \cite{Brain_Hubs-richclub2011}. Moreover,  their topological configuration develops over time from the prenatal to childhood and adult brain \cite{Brain_Hubs-development2015,Brain_Hubs-development2019}. The hubs also can play a crucial role in the appearance of diseases when their typical configuration becomes
  destroyed \cite{Brain_Hubs-diseases2012}.

Assuming that the higher-order connectivity may provide a clue
of how the hubs perform their function, here we examine the organisation of simplicial complexes around eight leading hubs in the human connectome. Based on our work \cite{Brain_weSciRep2019}, we use the consensus connectomes that we have generated at the Budapest connectome server \cite{http-Budapest-server3.0,Hung2}. These are connectomes that are common for one hundred female subjects (F-connectome) and similarly for one hundred male subjects (M-connectome), see Methods. Accordingly, we determine the hubs as eight top-ranking nodes in the whole connectome, performing the ranking according to the number of simplexes of all orders in which the node participates. These are the  Putamen, Caudate, Hippocampus and Thalamus-Proper in the left and similarly in the right brain hemisphere; they also appear as hubs according to several other graph-theory measures.  We then construct \textit{core networks} consisting of these hubs and all simplexes attached to them in both female and male connectomes. We determine the simplicial complexes and the related topological entropy in these core structures.  To highlight the weight-related heterogeneity of connections,  the structure of these core networks is gradually altered by increasing the threshold weight above which the connections are considered as significant.  We show that the
connectivity up to the 6th order remains in both connectomes
even at a high threshold. Meanwhile,  the identity of edges and their weights appear to be different in the F- and M-connectomes.

\section{Methods\label{sec:methods}}

\subsubsection{Data description}
 We use the data for two \textit{consensus connectomes}  that we have generated in  \cite{Brain_weSciRep2019} at the Budapest connectome server 3.0 \cite{http-Budapest-server3.0,Hung2} based on the brain imaging data from Human Connectome Project \cite{HCP_Neuroimage}. Specifically, these are the weighted whole-brain networks that are common for 100 female subjects, \textit{F-connectome}, and similarly, 
\textit{M-connectome}, which is common for 100 male subjects. Each connectome consists of $N=1015$ nodes annotated as the anatomical brain regions, and weighted edges, whose weight is given by the number of fibres between the considered pair of brain regions normalised by the average fibre length. Here, we consider the largest number 
$10^6$ fibres tracked and set the minimum weight to four. The
corresponding core networks \textit{Fc-network} and \textit{Mc-network} are defined as subgraphs of the F- and M-connectomes, respectively, containing the leading hubs and their first neighbour nodes as well as
all edges between these nodes. Meanwhile, the hubs are determined according to the topological dimension criteria, as described below and in Results. 

\subsubsection{Topology analysis and definition of quantities}
We apply the Bron-Kerbosch algorithm \cite{cliquecomplexes} to analyse
the structure of simplicial complexes, i.e., clique complexes, in the core Fc- and Mc-
connectomes. In this context, a \textit{simplex} of order $q$  is a
full graph (clique) of $q+1$ vertices $\sigma_q=\left \langle
  i_0,i_1,i_2,...,i_{q}\right \rangle$. Then a simplex $\sigma_r$  of the order
$r<q$ which consists of $r$ vertices of the simplex $\sigma_q$ is a
\textit{face} of the simplex $\sigma_q$.  Thus, the simplex $\sigma_q$
contains faces of all orders from $r=0$ (nodes), $r=1$ (edges), $r=2$ (triangles), $r=3$ (tetrahedrons), and so on,  up to the order $r=q-1$. A set of simplexes connected via shared faces of different orders makes a
\textit{simplicial complex}. The order of a simplicial complex is
given by the order of the largest clique in this complex, and
$q_{max}$ is the largest order of all simplicial complexes. 
Having the adjacency matrix of the graph, with the algorithm, we build
the incidence matrix ${\Lambda}$, which contain IDs of all
simplexes and IDs of nodes that make each simplex. With this
information at hand, we compute three structure vectors \cite{jj-book,Qanalysis1} to characterise the architecture of simplicial complexes:
\begin{itemize}
\item The \textit{first structure vector (FSV):}
$\mathbf{Q}=\{Q_0,Q_1,\cdots Q_{q_{max}-1}, Q_{q_{max}}\}$, where  $Q_q$
is the number of $q$-connected components;
 \item The \textit{second structure vector (SSV):} $\mathbf{N_s}=\{n_0,n_1, \cdots
n_{q_{max}-1},n_{q_{max}}\}$, where $n_q$ is  the number of
simplexes from the level $q$ upwards;
\item The \textit{third structure vector (TSV):}  the component $\hat{Q}_q \equiv
1-{Q_q}/{n_q}$ quantifies the degree of   connectedness
among simplexes \textit{at} the topology level $q$.
\end{itemize}
Furthermore, we determine the topological dimension of
nodes and topological entropy introduced in \cite{we-PRE2015}. The topological dimension $dimQ_i$ of a node $i$ is  defined
as the number of simplexes of all orders in which
the corresponding vertex participates, 
\begin{equation}
dimQ_i \equiv \sum_{q=0}^{q_{max}}Q_q^i \ ,
\label{eq:dimQ}
\end{equation}
 where $Q_q^i$  is determined directly from the
 ${\Lambda}$ matrix by tracking the orders of all simplexes in which the node
 $i$ has a nonzero entry.
Then, with this information, the entropy of a topological level $q$
defined as
\begin{equation}
S_Q(q) = -\frac{\sum_i p_q^i \log p_q^i}{\log M_q} \;
\label{eq-entropy-q}
\end{equation}
is computed. Here,  $p_q^i= \frac{Q_q^i}{\sum_i Q_q^i}$ is the node's 
occupation probability of the $q$-level, and the sum runs over all
nodes. The normalisation factor  $M_q=\sum_i\left(1-\delta_{Q_q^i,0}\right)$ is the number of vertices  having a nonzero entry at the level  $q$ in the entire graph.  
Thus the topological entropy (\ref{eq-entropy-q}) measures the degree
of cooperation among vertices resulting in a minimum at a given
topology level. Meanwhile, towards the limits $q\to 0$
and $q\to q_{max}$, the  occurrence of  independent cliques results in a higher entropy at that level.

In addition, we compute  the vector $\mathbf{f}= \left\{ f_0, f_1,
  \cdots  f_{q_{max}}\right\}$, which is defined 
\cite{we-PRE2015} such that $f_q$ represents  the \textit{number of simplxes and faces at the level} $q$. Given that a free simplex of the size
$n>q$ has the corresponding combinatorial number of faces of the order $q$, the component $f_q$ thus  contains information about the actual number of shared faces between simplexes \textit{at} the level $q$.

\subsubsection{Network structure \& hyperbolicity}
The underlying topological graph represents the 1-skeleton of the simplicial complex. Using the graph-theory methods \cite{sergey-lectures}, we determine the degree--degree correlations that are relevant to the observed "rich club behaviour" of the hubs in the global connectome \cite{Brain_weSciRep2019,Brain_Hubs-richclub2011,Brain_Hubs-development2015,Brain_Hubs-development2019}. 
Precisely, for each node in the considered network, the average number of edges of its nearest neighbour nodes is plotted against the node's degree. The
following scaling form is expected
\begin{equation}
\langle k\rangle_{i:nn} \sim k_i^\mu \ .
\label{eq:assort}
\end{equation}
Here, the positive values of the exponent $\mu > 0$ indicate the
\textit{assortative}  correlations, while $\mu < 0$ corresponds to a
\textit{disassortative} mixing, and $\mu=0$ suggests the absence of
nodes correlations. We analyse the Fc- and Mc-graphs by considering
the edges that remain after applying different weight thresholds. The weight distribution $P(w)$ is determined for the
entire core-networks, see Results.

Furthermore,  using the 4-point Gromov
criterion for the hyperbolic graphs \cite{HB-BermudoHBviasmallerGraph2013}, we determine the hyperbolicity parameter
$\delta_{max}$ of these graphs. Precisely, for each 4-tuple of nodes
$(A,B,C,D)$ in a $\delta$-hyperbolic
graph $G$,  the ordered relation between the sums of shortest-path distances 
${\cal{S}}\equiv d(A,B)+d(C,D) \leq {\cal{M}}\equiv  d(A,C) + d(B,D)
\leq {\cal{L}}\equiv d(A,D)+ d(B,C)$ implies that 
\begin{equation}
\delta(A,B,C,D) \equiv  \frac{{\cal{L}} - {\cal{M}}}{2} \leq \delta (G)\; .
\label{eq:hb-condition}
\end{equation}
It follows from  the triangle inequality that the upper bound of 
 $({\cal{L}}-{\cal{M}})/2$ is given by  the minimal distance
 $d_{min}\equiv min\{d(A,B),d(C,D)\} $ in the smallest sum
 ${\cal{S}}$. Thus, by sampling a large number
($10^9$)  4-tuples of nodes  in each graph, and plotting  $\delta(A,B,C,D) $
against the corresponding minimal distance $d_{min}$, we obtain $\delta (G)$
as the upper bound of $\delta_{max}=max_G\{\delta(A,B,C,D)\}$.

\section{Results\label{sec:results}}
\subsection{Whole-brain connectomes: Identification of hubs from
  topological dimension }
\begin{figure}[htb]
\centering
\begin{tabular}{cc} 
\resizebox{28pc}{!}{\includegraphics{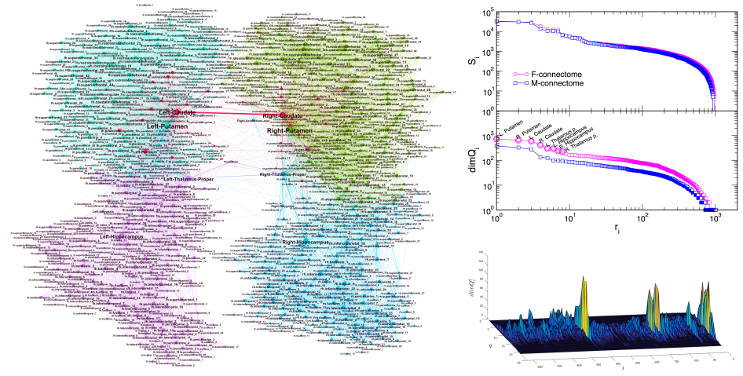}}\\
\end{tabular}
\caption{(left) F-connectome, 1000K fibres, with labels as brain
  areas.  (right, top) Ranking of the vertices according to the strength
  $S_i$, top, and topological dimension $dimQ_i$, lower panel, where eight leading vertices are marked (they also visible as hubs in the network on the left). (right,bottom) The 3D plot of the topological
  dimension against the topology level $q$ and the node's index
  $i$ for nodes in the F-core graph. }
\label{fig-ranking2x}
\end{figure}

We consider two whole-brain networks, precisely, the F-connectome,
which is common for 100 female subjects, and M-connectome,
consisting of the edges that are common to 100 maile subjects; see
Methods and \cite{Brain_weSciRep2019} for more details. For
illustration, the F-connectome is shown in the left panel of Figure\
\ref{fig-ranking2x}. Each connectome consists of 1015 nodes as
anatomical brain regions. These nodes are interconnected by a particular pattern of edges and organised in six mesoscopic communities.
 For this work, we determine the \textit{global hubs} in the F- and M-connectomes. These are eight top-ranking nodes according to the number of simplexes attached to a node. Based on our work in \cite{Brain_weSciRep2019}, we use the
 corresponding $\Lambda$-matrix for the F- and M-connectomes
 and identify simplexes of all orders in which a particular node
 $i=1,2,\cdots 1015$ participates. The node's topological dimension
 $dimQ_i$, defined by (\ref{eq:dimQ}) is then computed. For both connectomes, the node's ranking distribution by the decreasing topological dimension is shown in the middle right panel of figure\ \ref{fig-ranking2x}. As the figure demonstrates, the eight top-ranking nodes (marked along the curve for the F-connectome) make a separate group compared to the rest of the curve. These nodes also appear (see the list below)
 among the first eight ranked topological hubs in the M-connectome: 
\begin{verbatim}
rank_F     name                 rank_M 
1     Left Putamen              1
2     Right Putamen             3
3     Left Caudate              2
4     Right Caudate             4
5    Left Thalamus-Propper      5
6    Left Hippocampus           7
7    Right Hippocampus          8
8    Right Thalamus-Propper     6
\end{verbatim}
For comparisons with other approaches, we also show that these nodes
(with altered order) also appear as eight hubs ranked according to the node's strength $S_i$,  defined as the sum of weights of all edges of the node $i$. In this case, the ranking curves of the F- and M-connectomes
virtually overlap, see the top right panel in figure\
\ref{fig-ranking2x}. The lower right panel shows the 3-dimensional plot of the node's topological dimension over different topology
levels $q$. In this plot, the high peaks corresponding to our hubs
indicate what orders of simplexes mostly contribute to distinguishing the
hubs from the rest of the surrounding nodes.
Note that these eight nodes also appear as the leading hubs in several other sorting methods, for example, according to the node's degree and centrality measures \cite{Brain_Hubs-development2019,Brain_Hubs-richclub2011}.
For comparisons with other methods, we also show 
the names of nodes that rank from 9-20 according to the topological
dimension in the case of the F-connectome: 
\begin{verbatim} 
rh.precentral_7, Right-Pallidum, rh.caudalmiddlefrontal_11,
lh.caudalmiddlefrontal_13, Brain-Stem,  Left-Pallidum, 
lh.precentral_21, lh.precentral_6,lh.superiorparietal_25, 
rh.precentral_19, rh.precentral_15, rh.superiorparietal_13
\end{verbatim}
The nodes listed in the first two rows, except from the Brain Stem, also appear in this ranking range in the M-connectome.

Next, we consider a reduced network consisting of these
hubs and the nodes directly attached to any one of the hubs,
as well as the original edges between them in the  F- and
M-connectomes. The resulting \textit{core networks}
termed Fc- and Mc-networks, respectively, are shown in figure\ 
\ref{fig-coreNets2x}. Note that, by definition, the topological
dimension of the hubs is invariant to this network reduction.

\subsection{Core networks associated with  global
  hubs in the female and male connectomes}
\begin{figure}[htb]
\centering
\begin{tabular}{cc} 
\resizebox{28pc}{!}{\includegraphics{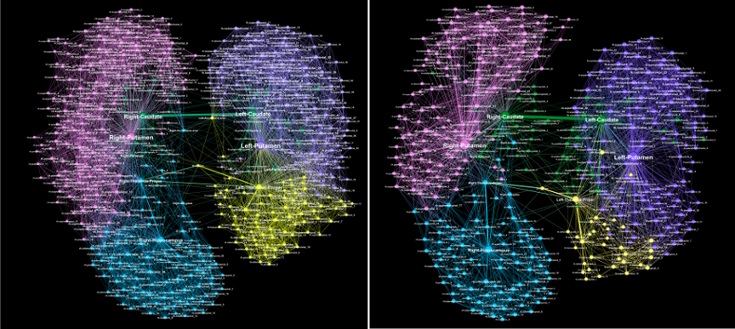}}\\
\end{tabular}
\caption{Core networks attached to the eight hubs in the female Fc- (left) and male
  Mc-connectome (right) from the original full-connectomes data at
  $N_F=10^6$ fibres tracked and  the  weight threshold $w_0=4$. The relative size of nodes is
  proportional to the number of their connections in the core-networks; 
  the node's labels show the corresponding anatomical brain region,
  and colours indicate five communities.}
\label{fig-coreNets2x}
\end{figure}

The extracted core Fc- and Mc-networks represent the part of the corresponding connectome in which the global hubs perform their function.  Here, we explore in detail the structure of the core networks in the female and male connectomes. Furthermore, we analyse how the structure depends on the weights of the edges. The histogram of the weights  is
shown in figure\ \ref{fig-PwFcMc}a for both Fc- and Mc-networks.
 As figure\ \ref{fig-coreNets2x} demonstrates, these core networks
exhibit a similar community structure. Precisely, each community in the
Fc- and similarly in Mc-connectome is a part of the global connectome
community, cf. figure\ \ref{fig-ranking2x}. This fact suggests that,  in both connectomes,  the core network reaches to all parts of the brain. Meanwhile,  it contains a smaller number of nodes (517 nodes in the Fc- and 418 in the Mc-network, respectively), and a considerably smaller number of connections compared to the whole connectome.  Thus, the node's assortativity changes as compared to the whole network. As the inset
to figure\ \ref{fig-PwFcMc} shows, the hubs mix in line with other
vertices in the core graphs, while they make a separate group when the
whole connectomes are considered \cite{Brain_weSciRep2019}.  This
assortative dependence emphasises the robustness of  core networks
with respect to  the hierarchical transmission of information among brain regions \cite{Nets_assortativegroups2018}.

\subsection{Topology of core networks depending on the weights
  of edges\label{sec:subnet}}
\begin{figure}[htb]
\centering
\begin{tabular}{cc} 
\resizebox{26pc}{!}{\includegraphics{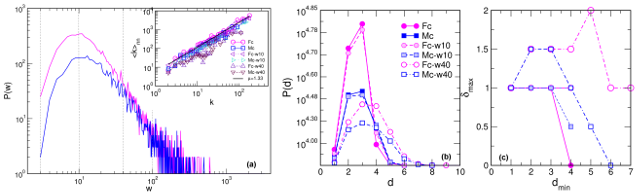}}\\
\end{tabular}
\caption{(a) Histogram of the weights of edges in the core Fc -and Mc-networks of the corresponding female and male connectomes, main
  panel; Inset: the assortativity plots of nodes in the core Fc- and
  Mc-networks for the weight threshold $w_0$=4, 10, 40, and 100,
  respectively, indicated by dotted vertical lines in the main panel.
  (b) Distribution of distances $P(d)$ against the shortest path
  distance $d$ and (c) the hyperbolicity parameter $\delta_{max}$
  against the shortest distance $d_{min}$ for the core Fc- and
  Mc-networks shown in figure  \ref{fig-coreNets2x}, and these networks for two larger threshold weights,  indicated in the legend.}
\label{fig-PwFcMc}
\end{figure}

\begin{figure}[htb]
\centering
\begin{tabular}{cc} 
\resizebox{24pc}{!}{\includegraphics{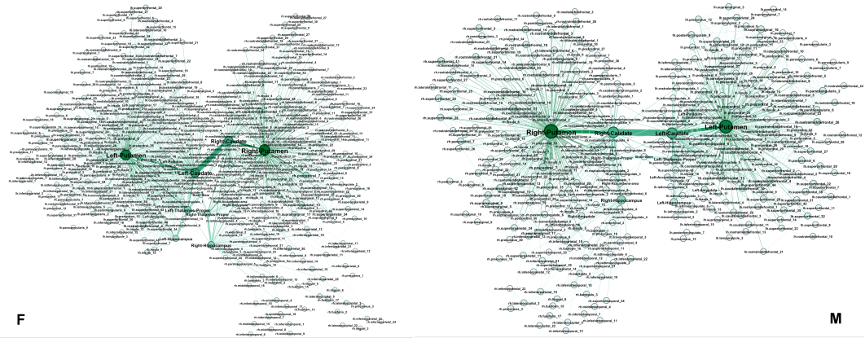}}\\
\end{tabular}
\caption{Core networks with the weights of edges above $w_0=40$ for the female  (left) and male (right) connectomes. Labels of nodes indicate the affected brain regions.}
\label{fig-coreNets2xw40}
\end{figure}

Using the approaches described in Methods, we determine several
algebraic-topology measures to characterise the structure of
simplicial complexes as well as the underlying topological graphs in the core Fc- and Mc-networks. 
These results are summarised in figures\ \ref{fig-PwFcMc} and \ref{fig-coreNets2xw40}.
Apart from a different number of nodes and edges that comprise the Fc-
and Mc-networks, we note that both of them are heterogeneous concerning the weight of edges, resulting in the broad log-normal
distributions in figure\ \ref{fig-PwFcMc}a.  Therefore, we obtain
different structures when the edges over a given threshold weight,  $w_0$, 
are considered. By gradually increasing the threshold $w_0=$10, 40, 100,
we show how the network properties change. More precisely, by removing the edges below the threshold, the network's diameter increases, and the distribution of the shortest-path distances change the
shape. Eventually, a larger cycle can appear, resulting in the
increased value of the hyperbolicity parameter, as shown in figure\
\ref{fig-PwFcMc}b,c. Meanwhile, the reduced networks preserve the
assortative mixing among the nodes, see the inset to figure \ref{fig-PwFcMc}a.

\begin{figure}[H]
\centering
\begin{tabular}{cc} 
\resizebox{26pc}{!}{\includegraphics{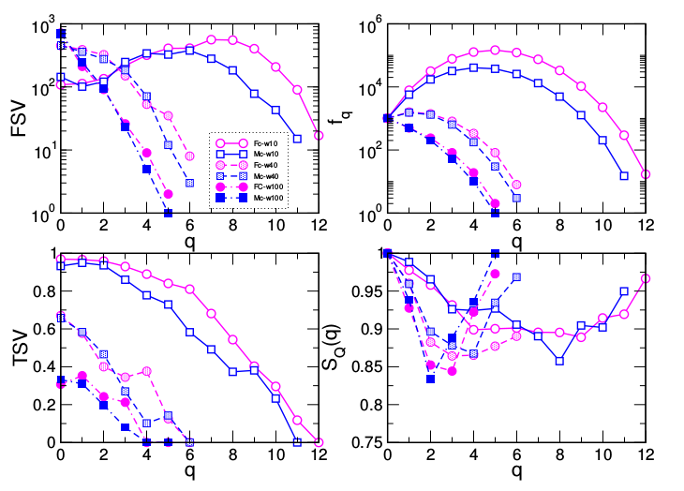}}\\
\end{tabular}
\caption{The first (FSV) and third (TSV) structure vectors,  the
  number of simplexes and faces $f_q$, and the topological entropy
  $S_Q(q)$  against the topology level $q$ in the core Fc- and Mc-networks with the edges of weights above the threshold $w_0=$10, 40, 100.
}
\label{fig-coreNets2xSVs}
\end{figure}
At the same time, the order of simplicial complexes gradually reduces
from $q_{max}=12$, in the case of $w_0=10$, to $q_{max}=5$ when edges
over the threshold $w_0=100$ are retained. 
The number of simplexes of the order $q=0,1,2\cdots q_{max}$,  given by
the FSV, and the ways that they interconnect, the TSV, change the
functional dependence of $q$ while at the same time reducing the difference between the Fc- and Mc-structures, cf.\ figure\ \ref{fig-coreNets2xw40}. The number of
simplexes and faces at the $q-$level, $f_q$, and the topological
entropy, $S_Q(q)$, follow a similar tendency. Moreover, the topological entropy
measure shows a pronounced minimum, indicating the
geometrical forms through which the nodes mostly interconnect. For example,
in the case of $w_0=100$, the minimum appears at $q=2$ (triangles) in
the Mc-, and $q=3$ (tetrahedrons) in the Fc-networks, respectively.
Figure \ref{fig-coreNets2xw40} illustrates the
remaining structures of the Fc- and Mc-networks when the weight
threshold $w_0=40$ is applied.

\begin{figure}[htb]
\centering
\begin{tabular}{cc} 
\resizebox{22pc}{!}{\includegraphics{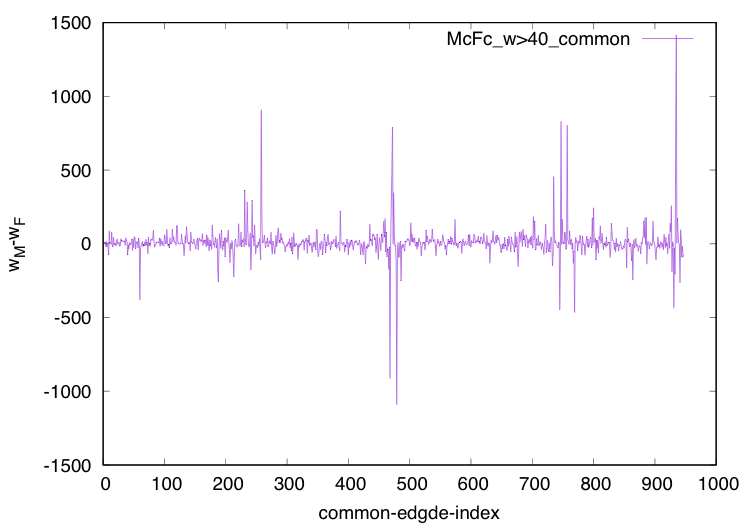}}\\
\end{tabular}
\caption{The weight difference $w_M-w_F$ of the common edges, indexed from 1 to 948,  in the Fc- and Mc-core networks in Fig.\
  \ref{fig-coreNets2xw40} with the edges weight over 40.}
\label{fig-mcfccommon-wdiff}
\end{figure}

An edge-to-edge comparison between the core Fc- and Mc-networks with
the threshold weight $w_0=40$, shown in  figure \ref{fig-coreNets2xw40}, 
revealed 948 edges that appear in both of them. Besides, the core
Mc-network has 204 unique edges that are not present in the
Fc-network with this threshold value, while the Fc-network has 419
such edges that are not seen in the corresponding Mc-network.
Moreover, the weight difference among the common edges varies, as
shown in figure  \ref{fig-mcfccommon-wdiff}. For example, the pairs of
nodes that make up 16 edges with a large difference $|w_M-wF| >300$  are listed below: 
\begin{verbatim}
ID  src_node              ID   dst_node         w_M   w_F
51 rh.parsopercularis_2  504 Right-Putamen      601   980
151 rh.precentral_9      504 Right-Putamen      740   375
159 rh.precentral_7      504 Right-Putamen     2453  1544
502 Right-Thalamus-Pr.  1008 Left-Thalamus-Pr.  809  1721
503 Right-Caudate       504 Right-Putamen      3340  2843
503 Right-Caudate       505 Right-Pallidum     3072  2280
503 Right-Caudate       507 Right-Hippocampus   937   591
503 Right-Caudate      1009 Left-Caudate       7122  8211
651 lh.precentral_21   1010 Left-Putamen       1220   765
654 lh.precentral_16   1008 Left-Thalamus-Pr.   137   584
654 lh.precentral_16   1010 Left-Putamen       1460   631
657 lh.precentral_4    1010 Left-Putamen       1796   993
661 lh.precentral_6    1010 Left-Putamen        618  1084
1008 Left-Thalamus-Pr. 1013 Left-Hippocampus   2481  2917
1009 Left-Caudate      1010 Left-Putamen       3174  2362
1009 Left-Caudate      1011 Left-Pallidum      3222  1805
\end{verbatim}

\section{Discussion and Conclusions\label{sec:discuss}}
We have analysed the structure of simplicial complexes surrounding eight topological hubs in the human connectomes. The hubs here
determined as the top-ranking nodes with the highest topological dimension (the number of simplexes attached). They represent the central
brain regions that coincide with the hubs determined by several other graph-theoretic measures. By parallel analysis of the
female and male consensus connectomes, we have extracted the
corresponding core segments, here termed the Fc- and Mc-networks, in
which the brain hubs perform their function.  Further, we have demonstrated that these
core networks are heterogeneous concerning the weights of
edges and they possess different weight-dependent
organisations. Consequently, their structure simplifies with the
increased weight threshold, eventually reducing at significant
thresholds to the 6-clique structure. Interestingly, these six nodes\begin{verbatim}
Right_Thalamus_Proper,Right_Caudate,Right_Putamen, 
Right_Pallidum, Right_Hippocampus, Right_Amygdala\end{verbatim} make up a remaining
6-clique structure in both female and male core networks. With two
additional nodes, we have found another 6-clique  in the female core
network, i.e.,
 \begin{verbatim}rh.precentral_15, rh.precentral_7, Right_Thalamus_Proper,
Right_Caudate,Right_Putamen, Right_Pallidum\end{verbatim} 
In both core networks, the identity of
the affected brain regions, as well as the variation of the weights
along the commonly present edges,  illustrates further differences between the female and male connectomes at the level of hubs.    

In the context of higher-order connectivity, these findings can
contribute to better understanding the pattern of connections that
enable the brain hubs to perform their role in the female and male
connectomes.
 Besides, the revealed detailed structure of simplicial complexes and the
identified brain regions that take part in them can facilitate the
desired simplex-based dynamics modelling of the brain functions.

%\section*{References}
%\bibliographystyle{unsrt}
%\bibliography{brainNets_17plus20}

\section*{\normalsize Acknowledgments}
Work supported  by  the Slovenian
Research Agency (research code funding number P1-0044). MA received
financial support from the Ministry of Education, Science and
Technological Development of the Republic of Serbia.  RM is also grateful for the NSERC and CRC programs for
their support.

\end{document}